\documentclass[12pt]{article} 
\usepackage{graphicx} 
\usepackage{dcolumn} 
\usepackage{bm} 
\usepackage[usenames,dvipsnames]{color}
\usepackage{CJK}
\usepackage{amsmath,amsfonts,amssymb}
\usepackage{fullpage}

\begin{document}


\title{The inexorable resistance of inertia determines the initial regime of drop coalescence}


\author{Joseph D. Paulsen, Justin C. Burton, Sidney R. Nagel \\
{\small The James Franck Institute and Department of Physics, The University of Chicago}\\ \\
Santosh Appathurai, Michael T. Harris, Osman A. Basaran \\
{\small School of Chemical Engineering, Purdue University}
}


\maketitle

\begin{abstract}
Drop coalescence is central to diverse processes involving dispersions of drops in industrial, engineering and scientific realms. During coalescence, two drops first touch and then merge as the liquid neck connecting them grows from initially microscopic scales to a size comparable to the drop diameters. The curvature of the interface is infinite at the point where the drops first make contact, and the flows that ensue as the two drops coalesce are intimately coupled to this singularity in the dynamics. Conventionally, this process has been thought to have just two dynamical regimes: a viscous and an inertial regime with a crossover region between them. We use experiments and simulations to reveal that a third regime, one that describes the initial dynamics of coalescence for all drop viscosities, has been missed. An argument based on force balance allows the construction of a new coalescence phase diagram.
\end{abstract}

The collision and coalescence of water drops, so essential to raindrop growth and the development of thunderstorms, have captivated the attention of the atmospheric science community since the early studies by Benjamin Franklin and Lord Rayleigh \cite{Sartor1969}.  Coalescence also plays a central role in industrial processes involving emulsions or dispersions \cite{Evans1994, Saboni1995}.  For example, in the petroleum industry, coalescence occurs during dispersed water removal and during oil desalting \cite{Eow2002}.  It is a dominant process in determining the shelf life of emulsion-based products such as salad dressing and mayonnaise \cite{Kumar1996}, and it occurs in dense spray systems and combustion \cite{Ashgriz1990}.  Also, sintering of two spherical particles closely resembles the coalescence of two dispersion drops in an emulsion \cite{Djohari2009}. Moreover, the controlled coalescence of drops in microfluidic devices promises a host of potential applications in chemistry, biochemistry, and materials science \cite{Ahn2006}.

The initial dynamics of coalescence are expected to be universal. The expansion of the liquid neck connecting two drops is controlled by the Laplace pressure, which diverges when the curvature of the liquid interface is infinite at the point where the drops first touch. Thus the change in topology, as two drops become one, is inextricably linked to a singularity in the dynamics.  Different regimes of coalescence have been studied \cite{Hopper1984,Hopper1990,Herrera1995,Eggers1999,MenchacaRocha2001,Eggers2003,Wu2004,Yao2005,Thoroddsen2005,Bonn2005,Lee2006,Burton2007,Fezzaa2008,Case2008,Case2009,Paulsen2011,Yokota2011}. The understanding that has emerged is that coalescence has just two dynamical regimes with a crossover region between them: a viscous regime, which always dominates at sufficiently early times when the neck radius is microscopically small, and an inertial regime that occurs at late times if viscous effects become negligible.  We use experiments and simulations to show that a third regime, one that describes the true initial dynamics of coalescence, has been missed.  We present our results in terms of a new phase diagram of coalescence that shows the three distinct dynamical regimes.\\

{\bf \large \noindent Results and Discussion}\\

\begin{figure*}
\centering
\includegraphics[width=16.5cm]{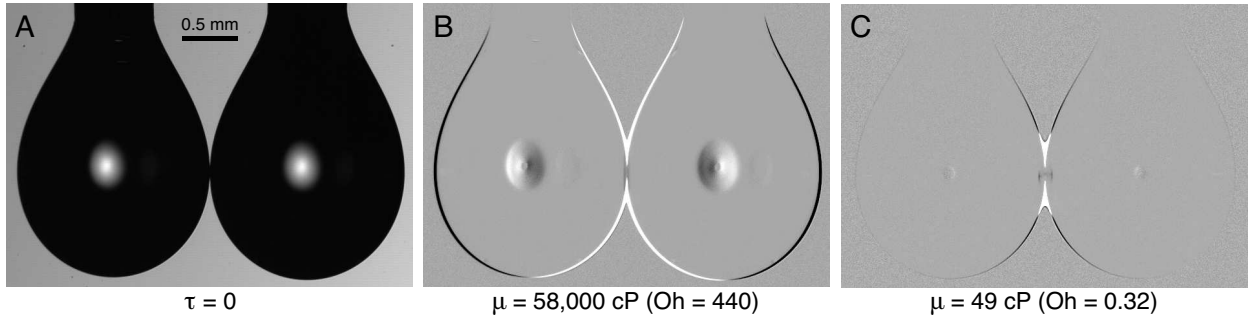} 
\caption{\footnotesize Coalescence of silicone oil drops with viscosities $\mu=$ 58,000 cP ($Oh=$ 440) and 49 cP ($Oh=$ 0.32). (A) Two pendant drops at the instant they contact, $\tau=$ 0. (The two bright spots are from back-lighting.) We subtract an image taken after the neck has grown to a size $r_{min}=0.25 A$ from the one at  $\tau=0$ for (B), $\mu=$ 58,000 cP and (C), 49 cP.}
\label{Fig1}
\end{figure*}

In the experiment, two pendant drops with radii $A\approx$ 0.1 cm are suspended as in Fig.\ \ref{Fig1}A from nozzles and slowly translated until they touch at their equators. Except where otherwise stated, the drops are silicone oil (surface tension $\gamma=$ 20 dyn/cm and density $\rho=$ 0.97 g/cm$^3$). The liquid viscosity, $\mu$, or equivalently the dimensionless Ohnesorge number, $Oh=\mu/\sqrt{\rho\gamma A}$, is varied. We use a high-speed camera and electrical resistance measurements \cite{Paulsen2011} to capture the dynamics. 

In the simulations, two isolated spherical drops of radius $A$ are connected by a small neck of radius $r_{min}=0.001A$. The dynamics that ensue are determined by solving the full Navier-Stokes equations by a finite-element algorithm that we have previously used successfully to study diverse situations involving drop breakup \cite{Chen2002,Suryo2006}. Creeping-flow simulations, where inertia is neglected, are also performed.

Fig.\ \ref{Fig1}A shows two drops at the instant of contact, $\tau=0$. To distinguish local versus global motion during merging, we subtract an image taken after the neck has grown to a small size from the one at $\tau=0$ as shown in Fig.\ \ref{Fig1}B and C for fluid viscosities $\mu=$ 58,000 cP ($Oh=440$) and 49 cP ($Oh=0.32$) respectively. At high viscosity, the two drops move together rigidly whereas at lower viscosity, the only appreciable motion occurs near the widening neck. This qualitative difference heralds the existence of a new regime.

The transition between these regimes can be understood from a force-balance argument that is based on the observation that in the perfectly viscous ({\em i.e.}, Stokes) regime, the drops outside the immediate vicinity of the neck are rigidly translated towards each other. This was shown by Hopper \cite{Hopper1984,Hopper1990} in an exact analytic solution of coalescence in two dimensions (2D); this global motion was also seen in 3D Stokes simulations \cite{Herrera1995} and the early-time asymptotic behavior was later analytically extended to three dimensions (3D) \cite{Eggers1999}. To be in the Stokes regime, therefore, the force of the neck pulling the two drops together must be sufficiently large to produce the required center-of-mass acceleration of each drop ({\em i.e.}, the fluid on each side of the $z=0$ plane). The asymptotic acceleration \cite{Hopper1984} is $a_{c.o.m.}=\gamma^2[\ln(r_{min}/8A)]^2/2\pi^2\mu^2 A$.

The coalescence is driven by surface tension. Therefore, an upper bound for the inward force of the neck on the drops is given by the surface-tension force, $F_{\gamma}$, around the circumference of the neck at its minimum radius. In 3D: $F_{\gamma}=2\pi\gamma r_{min}$. If $F_{\gamma}$ is too small to translate the drops, each having a mass $m=\frac{4}{3}\pi A^3\rho$, the flows cannot be in the Stokes regime. Therefore, the Stokes regime can only be achieved when $F_{\gamma}\gtrsim m a_{c.o.m.}$ leading to the threshold criterion for entering the Stokes regime:
\begin{eqnarray}
Oh \propto  \left|\ln\left(\frac{1}{8}\frac{r_{min}}{A}\right)\right|  \left(\frac{r_{min}}{A}\right)^{-1/2}  
\label{phaseboundary}
\end{eqnarray}

\noindent Therefore, for 3D drops of finite viscosity, the asymptotic dynamics, in the limit $r_{min}/A \rightarrow 0$, can {\em never} be in the Stokes regime. Below this threshold, $F_{\gamma}$ is balanced by the inertia of the drops, $ma_{c.o.m.}$, and the dynamics are governed by local flows. 

An analogous argument in 2D (where $F_{\gamma}=2\gamma$ independent of $r_{min}$, and $m=\pi A^2\rho$) suggests a phase-boundary for 2D drops: $Oh \propto |\ln(r_{min}/8A)|/\sqrt{4\pi}$. Thus the logarithmic divergence of $a_{c.o.m.}$ also precludes the Stokes regime in 2D until $r_{min}/A$ grows to a sufficient size.

\begin{figure}
\centering
\includegraphics[width=8.0cm]{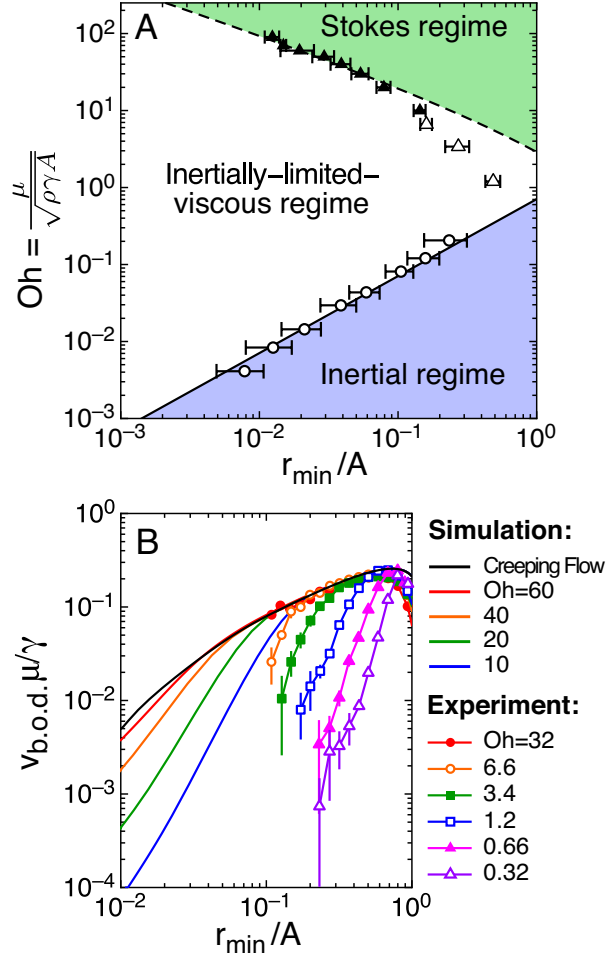} 
\caption{\footnotesize Phase diagram for 3D coalescence. (A) While the inertial regime \cite{Eggers1999,MenchacaRocha2001,Eggers2003,Wu2004,Thoroddsen2005,Bonn2005,Lee2006,Burton2007,Fezzaa2008,Case2008,Case2009,Paulsen2011} and the Stokes regime \cite{Hopper1984,Hopper1990,Herrera1995,Eggers1999,Yao2005} have been established in recent years, the inertially-limited-viscous regime is identified by this work. The inertially-limited-viscous to inertial crossover was recently determined by experiments (open circles) and a new scaling argument (solid line) \cite{Paulsen2011}. Here, we identify the inertially-limited-viscous to Stokes crossover with a force-balance argument (dashed line: eqn.\ \ref{phaseboundary} with a proportionality constant of 1.4), simulations (filled triangles) and experiments (open triangles). The data departs from the prediction at large $r_{min}$, where we expect finite size effects to enter. (B) To observe the inertially-limited-viscous to Stokes crossover, we measure $v_{b.o.d.}$ versus $r_{min}/A$ over a range of viscosities from simulation (solid lines) and experiment (symbols). The velocities fall onto the creeping-flow curve at large $r_{min}/A$, but peel away at smaller $r_{min}/A$. The velocity scaling at small $r_{min}/A$ is qualitatively captured by eqn.\ \ref{vInertialviscous}. The crossover neck radius at which the macroscopic drop velocity $v_{b.o.d.}$ merges onto the Stokes solution is plotted in (A), for the viscosities that exhibit such a crossover within our range of data.}
\label{Fig2}
\end{figure}

Fig.\ \ref{Fig2}A presents a phase diagram of the coalescence regimes for 3D drops, which includes the inertial and Stokes regimes in addition to this new ``inertially-limited-viscous" regime. In this regime, inertia and viscosity play a role in the dynamics; the inertia associated with each drop moving as a rigid object precludes the system from being in the Stokes regime. The inertially-limited-viscous to inertial crossover was previously determined \cite{Paulsen2011} to be $r_{min}/A \propto Oh$, in contrast to earlier work that had suggested that this crossover occurs at $r_{min}/A \propto Oh^2$ \cite{Eggers1999,Thoroddsen2005,Bonn2005}. (Previously, there were believed to be only two coalescence regimes---a viscous one and an inertial one---so this crossover is referred to as the viscous-to-inertial crossover in the literature.) The inertially-limited-viscous to Stokes transition is described by eqn.\ \ref{phaseboundary}. We emphasize that, contrary to earlier studies, we find the Stokes and inertial regimes do not share a phase boundary; they are both preceded by the inertially-limited-viscous regime. Thus, at early times, a model of pure Stokes flow for the coalescence of spheres is never valid. We note that this is reminiscent of the singularity in drop break-up \cite{Eggers2008}, where there are also three regimes, and the Stokes regime does not extend to $r_{min} \rightarrow 0$.

Having argued for the distinct identities of the inertially-limited-viscous regime versus the Stokes regime on theoretical grounds, we now offer evidence from experiment and simulation that these regimes are, in fact, different. First, we probe the global motion of the drops by measuring the velocity of the back of one drop, $v_{b.o.d.}$. Since this point is the farthest from the singularity, it isolates the global motion from the flow near the growing neck.

\begin{figure*}
\centering
\includegraphics[width=12.0cm]{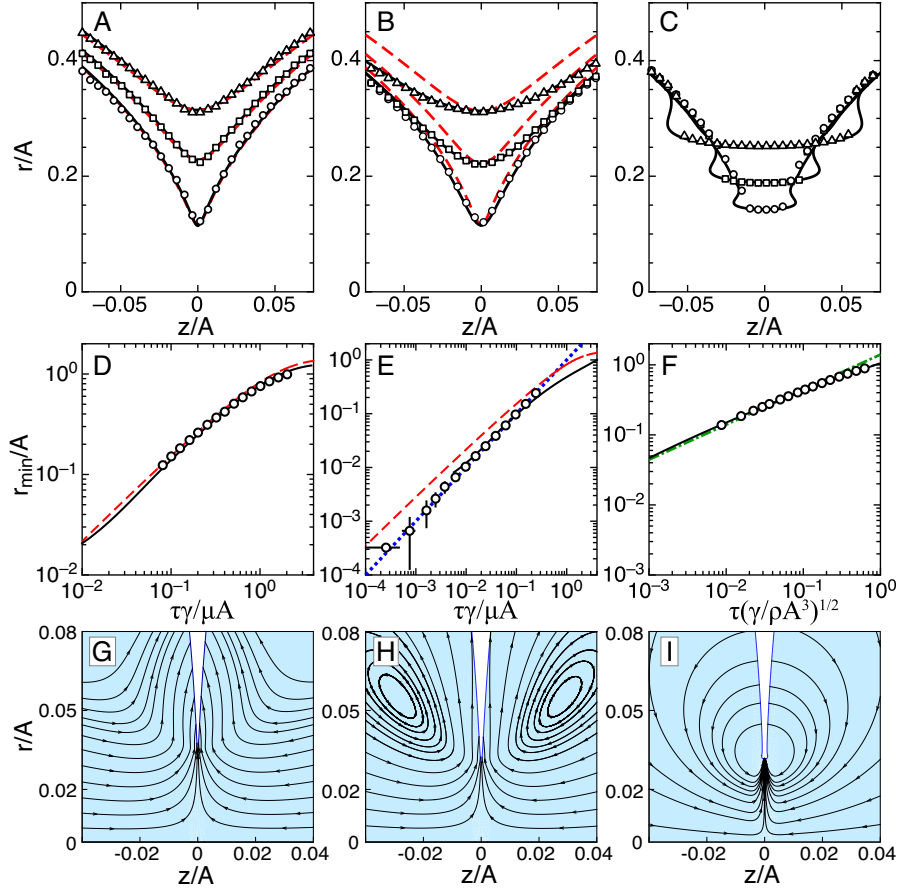} 
\caption{\footnotesize Coalescence dynamics in the Stokes regime (left column: creeping flow simulation, $Oh=440$ experiment) versus the inertially-limited-viscous regime (middle column: $Oh=0.6$ simulation and experiment) and the inertial regime (right column: $Oh=0.007$ simulation and experiment). (A,B,C) Neck profiles from simulation (solid lines) and experiment (symbols) at three different times, compared with the 2D Stokes theory in (A) and (B) (dashed lines \cite{Hopper1984}). The Stokes profiles agree with high-viscosity data (A) but do not capture the broader interfacial shapes at $Oh=0.6$ (B). (D,E,F) $r_{min}/A$ versus rescaled time in the simulation (solid lines) and experiment (circles). High-viscosity drops (D) follow with the Stokes theory (dashed line). At intermediate viscosity (E), $r_{min}/A$ does not agree with the Stokes theory but instead grows at the viscous-capillary velocity, $r_{min}=\tau\gamma/\mu$ (dotted line). At low viscosity in the inertial regime (F), $r_{min}/A=1.4 \tau^{1/2}(\gamma/\rho A^3)^{1/4}$ (dash-dot line). Note that in (E), we begin plotting the simulation data at $\tau\gamma/\mu A=6 \times 10^{-3}$, where we estimate that transients from the initial conditions have decayed. The experimental data in (E) was obtained by a high-speed electrical method on glycerol-salt-water drops ($\mu=$ 230 cP, $\gamma=$ 65 dyn/cm, $\rho=$ 1.2 g/cm$^3$, and $A=$ 0.2 cm) \cite{Paulsen2011}. (G,H,I) Instantaneous streamlines from simulations at $r_{min}=0.03 A$ in the (G) Stokes regime (creeping flow simulation), (H) the inertially-limited-viscous regime ($Oh=0.6$), and (I) the inertial regime ($Oh=0.007$). The flows are qualitatively different in all three regimes.}
\label{Fig3}
 \end{figure*}

For 3D drops in the inertially-limited-viscous regime, force-balance gives $a_{c.o.m.}\sim F_{\gamma}/m=3 \gamma r_{min}/2 A^3\rho$. Using $r_{min}=\tau\gamma/\mu$ as seen in Fig.\ \ref{Fig3}E consistent with previous experiments \cite{Bonn2005,Thoroddsen2005,Burton2007,Paulsen2011,Yokota2011}, we integrate to get:
\begin{eqnarray}
v_{b.o.d.} \sim \frac{3\gamma^2}{4\mu A^3\rho} \tau^2 = \frac{3 \mu}{4 A^3\rho} r_{min}^2.
\label{vInertialviscous}
\end{eqnarray}

\noindent If $Oh>1$, then the flows eventually enter the Stokes regime, where to first order:
\begin{eqnarray}
v_{b.o.d.} \approx \frac{\gamma}{2\pi\mu A} r_{min} \left|\ln \left( \frac{1}{8}\frac{r_{min}}{A} \right)\right|.
\label{vStokes}
\end{eqnarray}

We find that the creeping-flow simulation follows the exact analytic 2D Stokes solution (eqn.\ \ref{vStokes}). In Fig.\ \ref{Fig2}B, we plot $v_{b.o.d.}$ for drops of finite viscosity in the simulation and experiment. The curves exhibit the predicted super-linear growth of $v_{b.o.d.}$ at early times until the velocities merge onto the Stokes curve. The data show exceptional agreement between simulation and experiment. 

We measure the crossover neck radius at which the macroscopic drop velocity $v_{b.o.d.}$ merges onto the Stokes solution. We plot $Oh$ versus $r_{min}/A$ as the threshold for entering the Stokes regime on the phase diagram for 3D coalescence (Fig.\ \ref{Fig2}A). At higher viscosities, the linear (Stokes) regime is entered at smaller $r_{min}$. The data agree well with the prediction from eqn.\ \ref{phaseboundary}.

The neck shapes also differ between the Stokes and the intertially-limited-viscous regimes. Because the  exact analytic 2D solution extends over the entire domain of $r_{min}$ (whereas the 3D solution only exists for small $r_{min}$), we use it here as a convenient way to account approximately for finite-size effects and to compare our experiment and simulation. In Fig.\ \ref{Fig3}A and B, we plot this 2D Stokes solution \cite{Hopper1984} in the neck region against experiment and simulation, for $Oh\gg$ 1 and $Oh=$ 0.6, in the $r$-$z$ plane (with the origin at the initial point of contact). The Stokes solution agrees with the high-viscosity data, but it clearly fails to fit the shapes at $Oh=$ 0.6 where both experiment and simulation show a much broader neck. In particular, the 2D analytic Stokes solution has a maximum neck curvature, $\kappa$, that obeys $1/\kappa A=\frac{1}{4}(r_{min}/A)^3$ to first order, which we find is in good agreement with our data in the Stokes regime (i.e. $Oh\gg$ 1). In the inertially-limited-viscous regime (i.e. $Oh\sim$ 1), we also find that $1/\kappa A$ scales as $(r_{min}/A)^3$, but with a significantly larger prefactor (approximately 1.2 instead of $\frac{1}{4}$).

The dynamics also differ between these two regimes. For high-viscosity drops, measurements of the neck radius $r_{min}$ versus time in the experiment and simulation are consistent with the exact analytic 2D Stokes solution (Fig.\ \ref{Fig3}D).\footnote{Previous viscous coalescence experiments \cite{Yao2005,Thoroddsen2005,Bonn2005,Yokota2011,Paulsen2011} have compared the neck radius $r_{min}(\tau)$ against the theoretical prediction \cite{Eggers1999}, $r_{min}\approx \tau\gamma |\ln(\tau\gamma/\mu A)|/\pi\mu$. This approximate form breaks down \cite{Eggers1999} for $r_{min}>0.03 A$. Therefore, we compare our measurements against the full analytic 2D solution \cite{Hopper1984}, which can be done over the entire range of data.} For lower-viscosity drops, the 2D Stokes solution does not fit the data (Fig.\ \ref{Fig3}E). Instead, the neck radius grows linearly with time, consistent with $r_{min}=\tau\gamma/\mu$, a form that one might guess from dimensional analysis alone. (For the experimental data in Fig.\ \ref{Fig3}E, we coalesce hemispherical drops attached to circular nozzles separated by a distance $2A$. This altered boundary condition does not affect our results: using high-speed imaging, we find that for $r_{min}\ll A$, the dynamics is insensitive to this change in boundary conditions in both the inertially-limited-viscous and Stokes regimes.) This linear growth has been observed in previous experiments, but has incorrectly been assumed to be the dynamics of Stokes coalescence \cite{Bonn2005,Thoroddsen2005,Burton2007,Paulsen2011,Yokota2011}, and therefore was not recognized as evidence of a new regime. We emphasize that the observed power-law, $r_{min}\propto \tau$, is different from the inertial scaling where $r_{min}\propto \tau^{1/2}$ (as shown in Fig.\ \ref{Fig3}F and by previous work \cite{Eggers1999,MenchacaRocha2001,Eggers2003,Wu2004}), which demonstrates that the inertially-limited-viscous regime is distinct from the inertial regime as well.

Lastly, our simulations give the flow profiles near the singularity in the Stokes and the inertially-limited-viscous regimes (Fig.\ \ref{Fig3}G and H). The flow is expected to occur over a length scale comparable to $r_{min}$ in the Stokes regime \cite{Eggers1999}, and over a length scale $r_{min}^2/A$ in the inertially-limited-viscous regime \cite{Paulsen2011}. Indeed, while features in the creeping flow streamlines are roughly the size of $r_{min}$, the streamlines at intermediate $Oh$ exhibit recirculation zones, which constrict the flows near the neck. Comparing the geometry of the streamlines further solidifies that the new regime (Fig.\ \ref{Fig3}B,E,H)  is distinct from the one described by pure Stokes flow (Fig.\ \ref{Fig3}A,D,G) and the one described by inertial flow (Fig.\ \ref{Fig3}C,F,I). The streamlines in Fig.\ \ref{Fig3}I corroborate the definition of the Reynolds number proposed in \cite{Paulsen2011} that dictates the inertially-limited-viscous to inertial crossover.\\

{\bf \large \noindent Conclusion}\\

As two drops begin to coalesce and a microscopic liquid neck forms between them, the curvature of the interface and the Laplace pressure that develops due to surface tension both diverge at the instant when the drops first touch. In drop coalescence with no external fluid, previous work \cite{Eggers1999} incorrectly led to the conclusion that only viscous forces, along with surface tension, should dominate on small scales, a dynamical regime referred to as the Stokes regime. Our work identifies a necessary condition for Stokes flow to occur. The dynamics cannot be in the Stokes regime until the surface tension force around the neck is large enough to rigidly translate the two initially stationary drops towards each other. The inexorable resistance of inertia rears its head at even these small scales.

Therefore, a heretofore unknown dynamical regime controls the singularity at early times for drops of any viscosity. In this initial, asymptotic regime of drop coalescence in air, all of the underlying forces, that is inertial, viscous, and surface tension forces, are important. Hence, the two dynamical regimes referred to as the Stokes regime, where inertia is negligible, and the inertial regime, where viscous force is negligible, can only be attained once the neck has grown to a sufficient size.  Once the drop has entered the Stokes or inertial regimes, our measurements are consistent with the earlier predictions for the dynamics in those regimes.

A dynamically similar response is observed when a liquid filament breaks in air. At small neck size, the viscous and inertial regimes both give way to a third regime, where inertial, viscous and surface tension forces are all important \cite{Eggers2008}. When a liquid filament breaks in another liquid, however, the dynamics are qualitatively different: in that case the asymptotic dynamics of thinning may occur in the absence of inertia \cite{Lister1998, Cohen1999}. Further insight may likewise be gained by studying drop coalescence inside a second immiscible liquid, which is a problem of immense practical importance \cite{Leal2008, Cristini2001}.\\

{\bf \large \noindent Materials and Methods}\\

The experiments were conducted by authors JDP, JCB and SRN. High-speed imaging and electrical measurements were separately performed to capture the coalescence dynamics of isolated liquid drops in air. In high-speed imaging measurements, two silicone-oil drops with radii $A\approx$ 0.1 cm were suspended side by side as in Fig.\ \ref{Fig1}A from syringe needles. Different silicone oils were used in order to vary the liquid viscosity, $\mu$, from 49 cP to 58,000 cP, while keeping other fluid parameters constant (surface tension, $\gamma=$ 20 dyn/cm and density, $\rho=$ 0.96 g/cm$^3$ to 0.98 g/cm$^3$). Because they are highly wetting, pendant silicone-oil drops tend to climb up stainless steel syringe needles until the needles protrude from the bottoms of the drops. To prevent this, the needles were treated with an electronic coating (Novec$\texttrademark$ EGC-1700, 3M) that inhibits wetting. The drops were aligned and then slowly translated with a micrometer stage until they gently touched at their equators. The resulting coalescence dynamics were recorded with a high-speed digital camera (Phantom v12, Vision Research).

The electrical method is described in detail in \cite{Paulsen2011}. The experimental data in Fig.\ \ref{Fig3}E was obtained by this method on glycerol-salt-water drops ($\mu=$ 230 cP, $\gamma=$ 65 dyn/cm, $\rho=$ 1.2 g/cm$^3$, and $A=$ 0.2 cm).

The simulations were conducted by authors SA, MTH and OAB. The coalescence of two identical, isolated spherical drops of radius $A$ of an incompressible Newtonian fluid that are surrounded by a dynamically passive gas is simulated by connecting them with a small bridge of radius $r_{min}$ (typically equal to 0.1\% of the drop radius) and height $z_{min} \ll r_{min}$. The ensuing coalescence dynamics are governed by the continuity and the Navier-Stokes equations, i.e. the Navier-Stokes system. Because the two drops are identical and the two-drop configuration is axially symmetric, the computational domain is the planar quadrant that consists of one of the drops and one half of the bridge that is bounded by the plane of symmetry, the axis of symmetry, and the liquid-gas (L-G) interface. The Navier-Stokes system is solved subject to symmetry boundary conditions along the plane of symmetry and the axis of symmetry, and the kinematic and traction boundary conditions along the L-G interface \cite{Chen2002,Suryo2006}. This free boundary problem is solved numerically by a fully implicit method of lines (MOL) arbitrary Lagrangian-Eulerian (ALE) algorithm that uses the Galerkin/finite element method (G/FEM) for spatial discretization and an adaptive finite difference method (FDM) for time integration \cite{Chen2002,Suryo2006}.  On account of the free boundary nature of the problem, the interior of the flow domain is discretized by an adaptive elliptic mesh generation algorithm \cite{Christodoulou1992}.

The G/FEM converts the transient system of nonlinear partial differential equations (PDEs) to a system of nonlinear ordinary differential equations (ODEs). The FDM time integrator reduces the system of ODEs to a large system of nonlinear algebraic equations. This system of equations is then solved by Newton's method with an analytically calculated Jacobian.

Starting from an initially quiescent state, the dynamics are followed until the two drops have coalesced into one and the dynamics have ceased.  Simulations are carried out for both situations in which inertia is present, i.e. $Oh$ is finite, and also when inertia is negligible, i.e $1/Oh = 0$, such that the drops undergo creeping (Stokes) flow.\\

{\bf \large \noindent Acknowledgements}\\

We thank Michelle Driscoll, Efi Efrati and Wendy Zhang. This work was supported by NSF grant No. DMR-1105145, the University of Chicago NSF-MRSEC DMR-0820054, the NSF ERC-SOPS (EEC-0540855), and the BES Program of the US DOE. Use of facilities of  the Keck Initiative for Ultrafast Imaging is gratefully acknowledged.

\bibliographystyle{hplain}
\bibliography{CoalescenceNewRegimeBib}

\end{document}